\documentclass[fleqn,usenatbib]{mnras}

\usepackage{newtxtext,newtxmath}

\usepackage[T1]{fontenc}

\DeclareRobustCommand{\VAN}[3]{#2}
\let\VANthebibliography\thebibliography
\def\thebibliography{\DeclareRobustCommand{\VAN}[3]{##3}\VANthebibliography}

\usepackage{graphicx}	
\usepackage{amsmath}	

\usepackage{threeparttable}
\usepackage{caption}

\title[Mass constraints of AMXPs]{Accretion-induced spin-up: Implications for mass constraints of AMXPs}

\author[X. Y. Lai]{
Xiaoyu Lai,$^{1}$\thanks{E-mail: laixy@hue.edu.cn (XYL)}
\\
$^{1}$Institute of Astronomy and High Energy Physics, Hubei University of Education, Wuhan 430205, China
}

\date{Accepted XXX. Received YYY; in original form ZZZ}

\pubyear{\the\year{}}

\begin{document}
\label{firstpage}
\pagerange{\pageref{firstpage}--\pageref{lastpage}}
\maketitle

\begin{abstract}
We investigate the influence of the global structure of accreting millisecond X-ray pulsars (AMXPs) on accretion-induced spin-up, using three equations of state (EoS) models representing neutron stars, quark stars, and strangeon stars. By applying the classical accretion torque formalism, and deriving the accretion rate and magnetic field from observations of three AMXPs --- XTE J1751-305, SAX J1808.4-3658, and IGR J00291+5934 --- we derive mass contours in the plane of luminosity versus spin-up rate. Our results show that the inferred masses are broadly consistent across the three EoS models, indicating that the method is insensitive to the specific EoS and can serve as an evolutionary channel for constraining the masses of pulsar-like compact stars. Notably, we find that SAX J1808.4-3658 and IGR J00291+5934 are constrained to very low masses, while XTE J1751-305 yields a mass consistent with the typical range for pulsars. 
Our analysis suggests that future improvements in distance and spin-up measurements would refine these mass constraints, which could offer crucial evidence to distinguish different EoS models.
This paper also presents the idea that using EoS-specific parameters could yield new insights.
\end{abstract}

\begin{keywords}
keyword1 -- keyword2 -- keyword3
\end{keywords}



\section{Introduction}

Born in supernova explosions, pulsar-like compact stars have long been studied for their internal structure. 
Observables that reflect their global properties, such as mass and radius, offer a standard route to constraining the equation of state (EoS). 
Additionally, the environments hosting pulsars are often highly diverse, and the evolution of pulsars within these environments manifests in a variety of observational features. 
If evolutionary pathways that depend on stellar structure can be recognized, it may become possible to use observational data to infer structural properties.

Accreting millisecond X-ray pulsars (AMXPs) are neutron star low-mass X-ray binaries (LMXBs) that exhibit coherent X-ray pulsations, powered by plasma accreted from the disc. 
Since the first discovery of such pulsations in SAX J1808.4-3658~\citep{Nature1998}, about two dozen AMXPs have been detected~\citep[see Table 1 in][for an overview]{2019AA}. 
These systems are thought to represent the evolutionary link between neutron star LMXBs containing and spin-powered millisecond pulsars (MSPs).


The X-ray outbursts of AMXPs, triggered by a sudden increase in the mass accretion rate, are generally accompanied by spin-up. 
However, some AMXPs have been observed to exhibit spin-down during outbursts (e.g., IGR J17591-2342~\citep{IGR.J17591.2020MNRAS}), indicating that the accretion torque could be highly complex due to interactions between the accretion disc and the magnetic field. 
In an LMXB, the spin evolution of the neutron star, for a given accretion rate and magnetic field strength, is governed by its global structure, which in turn is determined by the EoS of ultra-dense matter~\citep[see Section 4.1 in][]{Salvo2022}. 
Hence, accretion-induced spin-up could, in theory, serve as a viable channel for probing the EoS.

Several factors, however, necessitate a careful selection of the sample when investigating accretion-induced spin evolution.
The pulse phase evolution during outbursts often exhibits strong timing noise~\citep{Patruno2009ApJ}, which can hamper the clean measurement of the spin derivative. 
Moreover, the intrinsic X-ray luminosity is crucial for us to derive the accretion rate, but in most cases it is uncertain, mainly due to the uncertainty in the distance.
Therefore, only a limited number of sources are currently available for investigating this issue~\citep{Patruno2021ASSL}.

For a given EoS model, the values of $M$, $R$ and $I$ are uniquely determined by the central density, so any one of these quantities --- for instance, $M$ --- can be used to represent the global stellar structure. 
On the other hand, besides the accretion rate and magnetic field, 
the efficiency of accretion-induced spin-up also depends on $M$, $R$, and $I$. 
Therefore, accretion-induced spin-up could, in principle, serve as an evolutionary channel for testing EoS models or constraining the values of $M$. 
This method, which we apply in this paper, differs substantially from the often used approaches that adopt fixed typical values of $M$, $R$ and $I$ when studying accretion-driven spin evolution.

While adopting the typical structural parameters may not lead to substantial errors, alternative approaches also deserve consideration. 
Several studies have investigated how particular EoS models shape the outcomes of astrophysical processes.
For example, the constraints on radius of SAX J1808.4–3658 using some EoS models have been made~\citep{Burderi1998}.
Another related example is the investigation on how different EoS models affect the minimum attainable spin period of MSPs~\citep{ZhongMNRAS2025}.
It would be worthwhile to extend this approach to more processes.

In this paper, we use three EoS models, including AP model~\citep{AP1997} for neutron stars (NSs), MIT bag model~\citep[see][and references therein]{Bhattacharyya2016MNRAS} for quark stars (QSs) and Lennard-Jones model~\citep{LX2009MNRAS} for strangeon stars (SSs).
We choose three AMXPs which are thought to have the best-measured spin evolution during and after the outbursts.
In Section~\ref{sec:sec_method} we investigate how the global structure of AMXPs affects the
accretion-induced spin-up, and the results will show that there are not much differences among different EoS models. 
This may indicate that accretion-induced spin-up could theoretically serve as an evolutionary channel to constrain the masses of AMXPs.
Notably, we find that SAX J1808.4-3658 and IGR J00291+5934 are inferred to have very low masses.
Conclusions and discussions are made in Section~\ref{sec:sec_con}.

\section{The effect of stellar structure on spin-up}
\label{sec:sec_method}

\subsection{Selected sources}
\label{sec:sources}

To investigate how the global structure of AMXPs affects the accretion-induced spin-up, in this paper We select one outburst from each of three sources showing recurrent outbursts, XTE J1751-305, SAX J1808.4-3658 and IGR J00291+5934.
They were also selected by~\citet{GS.MNRAS2021} to investigate how the uncertain physics in the vicinity of the magnetospheric radius affects the spin-up episodes in AMXPs.
These sources have currently the best-measured values of spin evolution during and after the outbursts~\citep{GP.J1751,Burderi2006MNRAS,Falanga2005A&A}, as well as stable long-term spin-down during quiescent phases, which could be used to estimate the polar magnetic filed strength~\citep{Patruno2021ASSL}.

The observed properties are listed in Table~\ref{tab:sources}, including the spin frequency $\nu$, the time derivative of spin frequency during outburst $\dot{\nu}_{\rm su}$ and quiescent phase $\dot{\nu}_{\rm sd}$, and the bolometric X-ray luminosity $L_{\rm X}$. 
For the latter three quantities, we use their mean values ($\left<\dot{\nu}_{\rm su}\right>$, $\left<\dot{\nu}_{\rm sd}\right>$ and $\left<L_{\rm X}\right>$) which can be reliably measured, as the observables.

\begin{table*}
	\centering
    \begin{threeparttable}
	\caption{Observed properties of three outbursts of the three AMXPs considered in this work.}
	\label{tab:sources}
	\begin{tabular}{lcccr} 
		\hline
		Source & $\nu$ (Hz) & $\left<\dot{\nu}_{\rm su}\right>$ (Hz s$^{-1}$) & $\left<\dot{\nu}_{\rm sd}\right>$ (Hz s$^{-1}$) & $\left<L_{\rm X}\right>$ ($\times \rm\ 10^{38}\ erg\ s^{-1}$)\\
		\hline
		XTE J1751-305 (2002)$^{\rm a}$ & 435.3 & $3.7(10)\times 10^{-13}$ & $-5.5(12)\times 10^{-15}$ & $\sim 0.25\times (d/8.5{\rm\ kpc})^2$\\
		SAX J1808.4-3658 (2002)$^{\rm b}$ & 401.0 & $4.40(81)\times 10^{-13}$ & $-5.6(20)\times 10^{-16}$ & $\sim 0.1\times (d/3.5{\rm\ kpc})^2$\\
		IGR J00291+5934 (2004)$^{\rm c}$ & 598.9 & $5.1(3)\times 10^{-13}$ & $-4.1(12)\times 10^{-15}$ & $\sim 0.063\times (d/5{\rm\ kpc})^2$\\
		\hline
	\end{tabular}
    \begin{tablenotes}
            \footnotesize  
            \item[a] \citet{Papitto2008MNRAS}; \citet{Riggio2011A&A}; \citet{GP.J1751}
            \item[b] \citet{Burderi2006MNRAS}; \citet{Hartman2008ApJ}
            \item[c] \citet{Patruno2010ApJ}; \citet{Papitto2011A&A}; \citet{Falanga2005A&A}
        \end{tablenotes}
    \end{threeparttable}
\end{table*}

\subsection{Accretion torque}
\label{sec:torque}

Although some more detailed modeling of spin-up torque have been studied, e.g. by~\citet{GS.MNRAS2021}, to concentrate on the influence of the stellar structure on the accretion-induced spin-up, we will use the classical form of accretion torque. 

A pulsar spinning at angular velocity $\Omega$ and accreting with rate $\dot{M}$ will be exerted a torque written as~\citep{Menou1999ApJ,Chatterjee2000ApJ}
\begin{equation}
\dot{J}=2\dot{M}R_{\rm m}^2\Omega_{\rm K}(R_{\rm m})\left(1-\frac{\Omega}{\Omega_{\rm K}(R_{\rm m})}\right), \label{eq:torque}
\end{equation}
where 
\begin{equation}
R_{\rm m}\simeq[B^2R^6/(\dot{M}\sqrt{2GM})]^{2/7} \label{eq:Rm}
\end{equation}
is the magnetospheric radius defined as the location inside which the magnetic energy density of the falling material will overwhelm the kinetic energy, and  
\begin{equation}
\Omega_{\rm K}(R_{\rm m})=\sqrt{\frac{GM}{R_{\rm m}^3}} \label{eq:Omega_K}
\end{equation}
is the Keplerian angular velocity at radius $R_{\rm m}$.
Here $M$, $R$ and $B$ are the mass, radius and surface dipole magnetic field of the pulsar, respectively.

The torque in Eq.(\ref{eq:torque}) is related to the derivative of spin angular velocity $\dot{\Omega}$ via 
\begin{equation}
\dot{J}=I\dot{\Omega},\label{eq:Jdot}
\end{equation}
where $I$ is the moment of inertia of the pulsar.

\subsection{Accretion rate}
\label{sec:Mdot}

Assume that all the gravitational energy released during accretion is radiated via X-ray, we can get the canonical relation between the accretion rate $\dot{M}$ and the bolometric X-ray luminosity $L_{\rm X}$,
\begin{equation}
L_{\rm X}\simeq\frac{GM\dot{M}}{R}, \label{eq:Lx}
\end{equation}
which shows that, the accretion rate $\dot{M}$ derived from the observed X-ray luminosity $L_{\rm X}$ also depends on $M$ and $R$.

The uncertain about the distance to a source, however, gives uncertainty about the luminosity deduced from the observed flux.
In practice, when considering the accretion-induced spin evolution, the often used method is extracting $\dot{M}$ from the accretion torque, in which the typical values $M$, $R$ and $I$ are used. 
Then $\dot{M}$ can be used to constrain the distance to the source, and then the corresponding luminosity.
For the three systems mentioned above, the related work can be found in~\citet{Papitto2008MNRAS} and ~\citet{Burderi2006MNRAS,Burderi2007ApJ}.
Since our aim is to study the influence of stellar structure on accretion-induced spin-up, we should not use the luminosity derived in such way. 

A good estimation of luminosity is from the bursts exhibiting the photospheric radius expansion (PRE) which can be used as standard candles~\citep{Kuulkers2003A&A}.
Using this method, the distance of SAX J1808.4-3658 is estimated to be 2.5-3.6 kpc from the outbursts in 1996 and 2002~\citep{intZand2001A&A,Galloway2006ApJ}, and the distance of IGR J00291+5934 is estimated to be 4.2$\pm$0.5 kpc from the outburst in 2015~\citep{DeFalco2017A&A}.
However, XTE J1751–305 shows no such kind of outbursts so its distance is difficult to determine, and the upper limit is often set to be 8.5 kpc if one assumes that the source is close to the Galactic center.

Using the distances estimated above, the luminosities shown in Table~\ref{tab:sources} can be re-evaluated accordingly, then we can get $\dot{M}$ by Eq.(\ref{eq:Lx}) and left $M$ and $R$ as the variables to be determined.

\subsection{Magnetic field}
\label{sec:B}

For AMXPs showing recurrent outbursts, the long-term spin evolution can be estimated by comparing the averaged spin frequency measured in each outburst. 
Although the long-term spin-down could be attributed to three mechanisms~\citep{Hartman2008ApJ,Papitto2011A&A}, including the magnetic dipole radiation, the propeller torque and the gravitational radiation, here we neglect latter two mechanisms.
This could be reasonable, since the three sources we selected in this paper have stable long-term spin-down rates, suggesting that its origin is more likely to be dominated by the magnetic dipole radiation~\cite{Patruno2021ASSL}. 
This also means that the magnetic threading of the accretion disk outside the corotation radius~\citep{Wang1995ApJ,Rappaport2004ApJ}, which will give a braking torque different from the one given by the magnetic dipole field of the pulsar, will not be considered here.

If the long-term spin-down during quiescent phases is dominated by the magnetic dipole radiation, the surface dipole magnetic field could be derived by~\citep{Spitkovsky2006ApJ}
\begin{equation}
B=\sqrt{\frac{c^3IP\dot{P}}{4\pi^2R^6}\frac{1}{1+\sin^2\alpha}}, \label{eq:B}
\end{equation}
where $P$ is the spin period ($P=1/\nu$), $c$ is the speed of light, and $\alpha$ is the angle between the magnetic and rotational poles.
The time derivative of the spin period $\dot{P}=-{\dot{\nu}_{\rm sd}}/\nu^2$, where values of $\nu$ and $\left<{\dot{\nu}_{\rm sd}}\right>$ are listed in Table~\ref{tab:sources}.
Here we use the mean values $\left<{\dot{\nu}_{\rm sd}}\right>$ and ignore the errors.
The difference in $B$ brought by different $\alpha$ will not lead to qualitatively different results, so in the following calculations we assume that $\sin \alpha=0.5$.

\subsection{Results}
\label{sec:results}

From Eqs.(\ref{eq:torque}) to (\ref{eq:B}), if we know $\nu$, $\dot{\nu}_{\rm sd}$ and $L_{\rm X}$ from observations, as well as $M$, $R$ and $I$ from a given EoS model, we can derive the value of $\dot{\nu}_{\rm su}$. 
Because the values of $M$, $R$ and $I$ are uniquely determined by a given central density under an EoS model, we can eventually determine the value of $\dot{\nu}_{\rm su}$ for a fixed value of $M$.
The contours of $M$ can then be plotted in the $L_{\rm X}$-$\dot{\nu}_{\rm su}$ plan.

For an AMXPs, X-ray luminosity and spin-up rate can vary from outburst to outburst, but the spin frequency and the long-term spin-down rate in quiescence are stable. 
Therefore, contours of $M$ under a certain EoS model plotted in the $L_{\rm X}$-$\dot{\nu}_{\rm su}$ plan for a given AMXP are fixed, and $M$ can be constrained by comparing the observed points ($L_{\rm X}$ and $\dot{\nu}_{\rm su}$) with the contours.

\begin{figure}
	\centering
    \includegraphics[width=0.9\columnwidth]{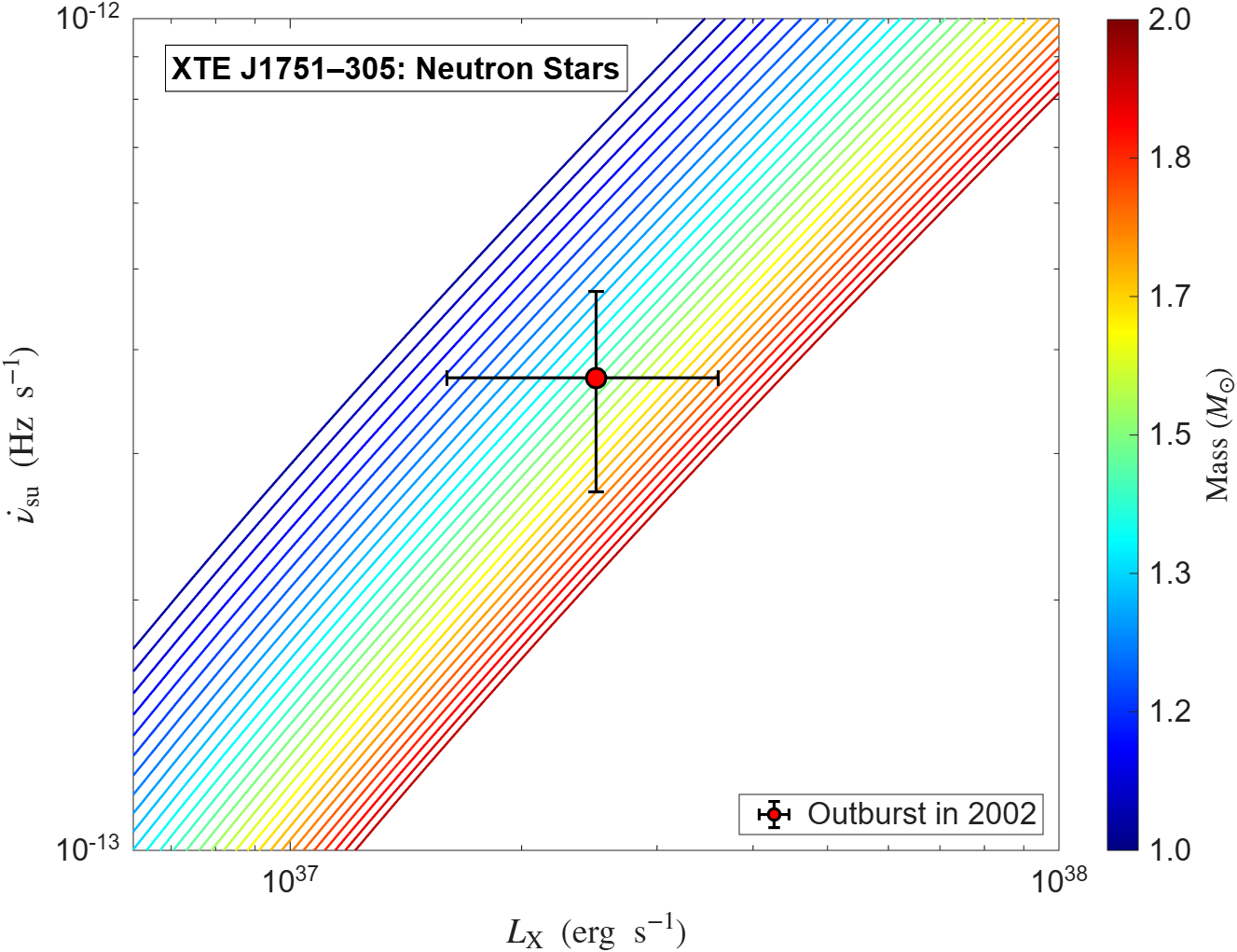}
    \includegraphics[width=0.9\columnwidth]{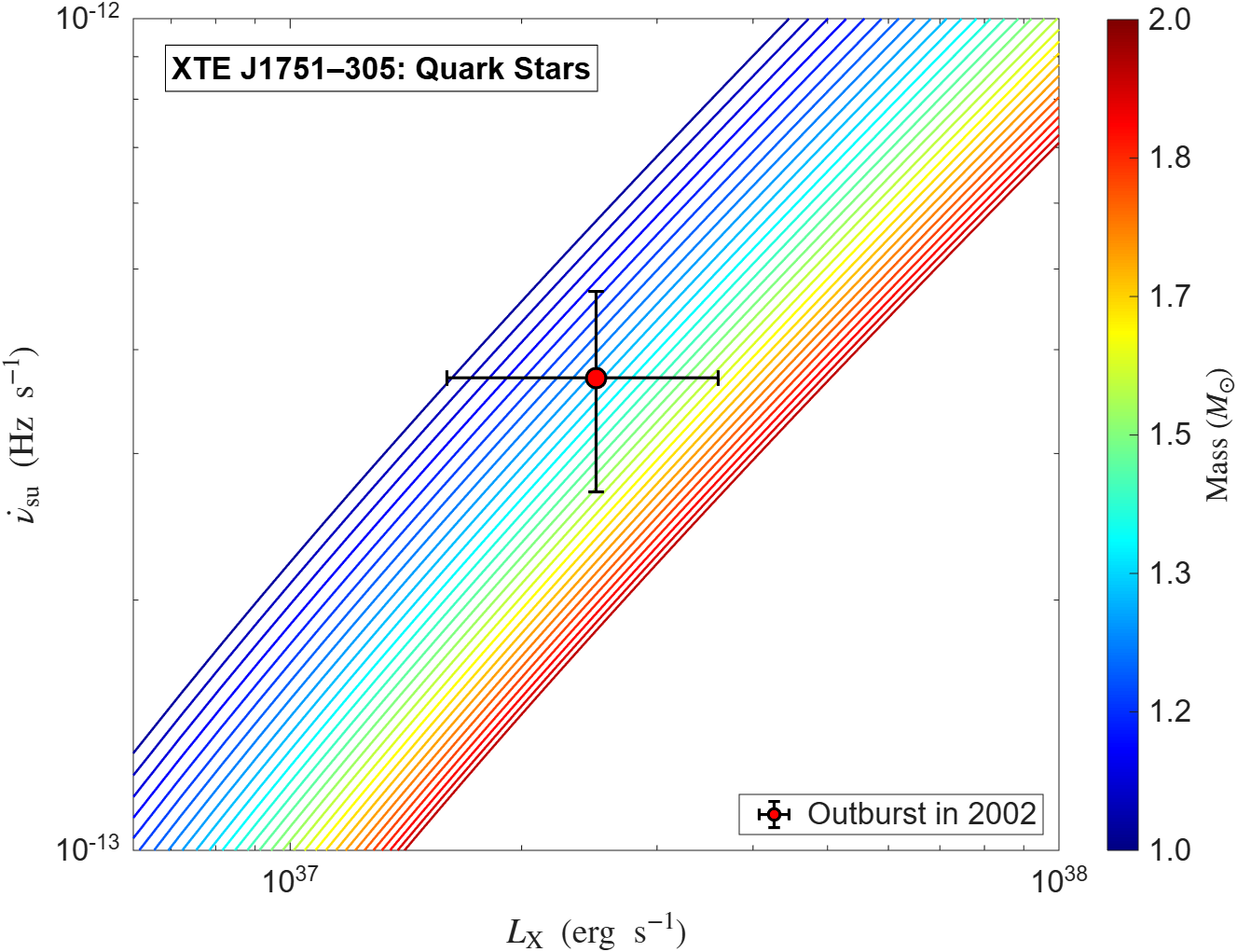}
    \includegraphics[width=0.9\columnwidth]{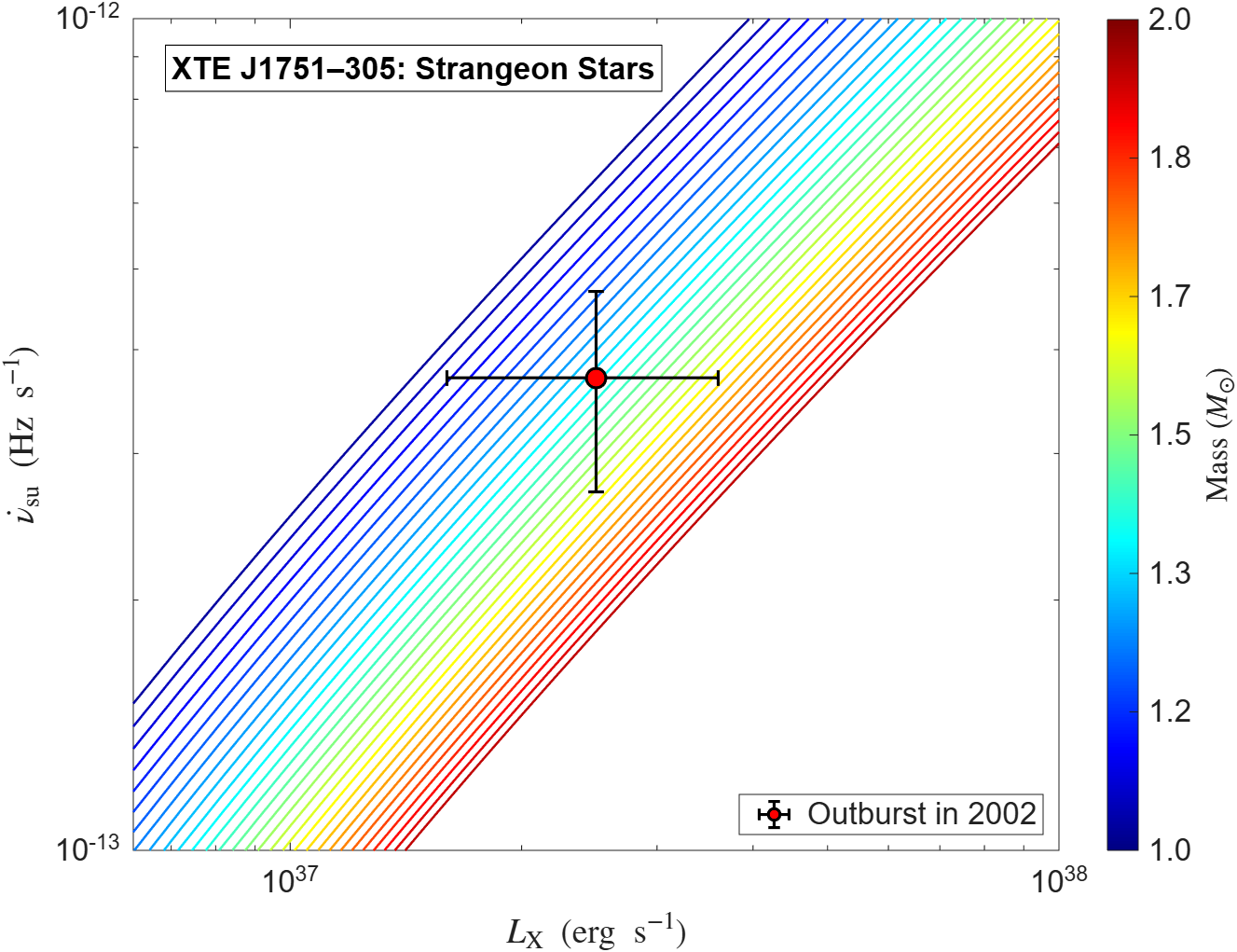}
    \caption{Contours of $M$ with different colors (upper purple one to lower red one: 1 to 2 $M_\odot$) for XTE J1751–305, with top, middle and bottom panels representing the cases of neutron stars, quark stars and strangeon stars, respectively. The data point is from the outburst in 2002. The errors in $\dot\nu_{\rm su}$ are from the third column of Table.~\ref{tab:sources}. Because the distance of XTE J1751–305 is uncertain and is often taken to be 8.5 kpc, the errors in $L_{\rm X}$ are given assuming that the error in the distance is 20 percent.   
    }
    \label{fig:results_1751}
\end{figure}

The results for XTE J1751–305 are shown in Fig. \ref{fig:results_1751}, with top, middle and bottom panels representing the cases of neutron stars, quark stars and strangeon stars, respectively.
In each panel, the contours of $M$ with different colors correspond to masses ranging from 1 to 2 $M_\odot$, from upper purple one to lower red one.
The data point is from the outburst in 2002. 
The errors in $\dot\nu_{\rm su}$ are from the third column of Table.~\ref{tab:sources}.
Because the distance of XTE J1751–305 is uncertain, we take its often used value 8.5 kpc~\citep{GP.J1751} and give the errors in $L_{\rm X}$ assuming that the error in the distance is 20 percent. 
It can be seen that, although the three EoS models yield slightly different results, the data point falls within a similar range around 1.4 $M_\odot$.

\begin{figure}
	\centering
    \includegraphics[width=0.9\columnwidth]{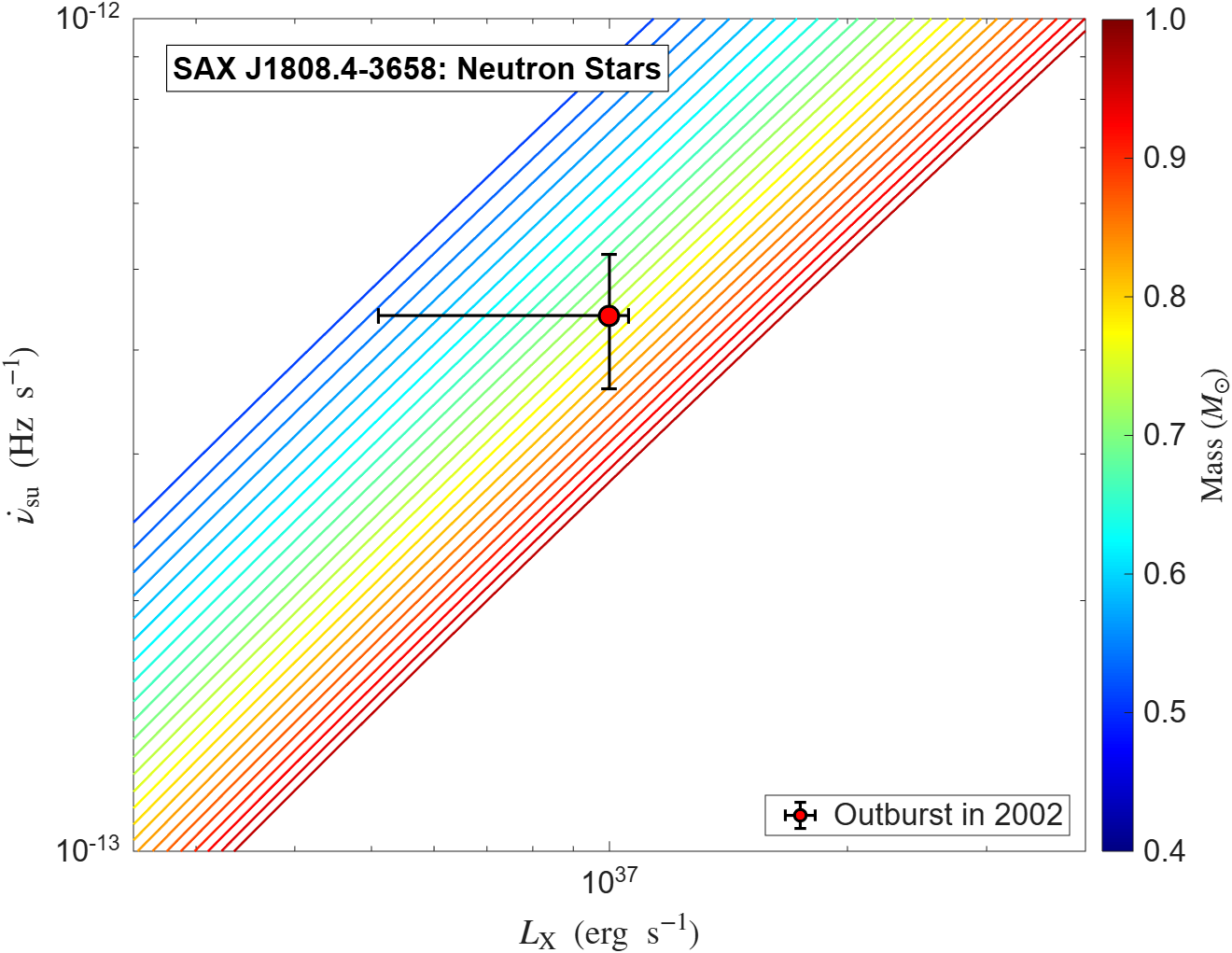}
    \includegraphics[width=0.9\columnwidth]{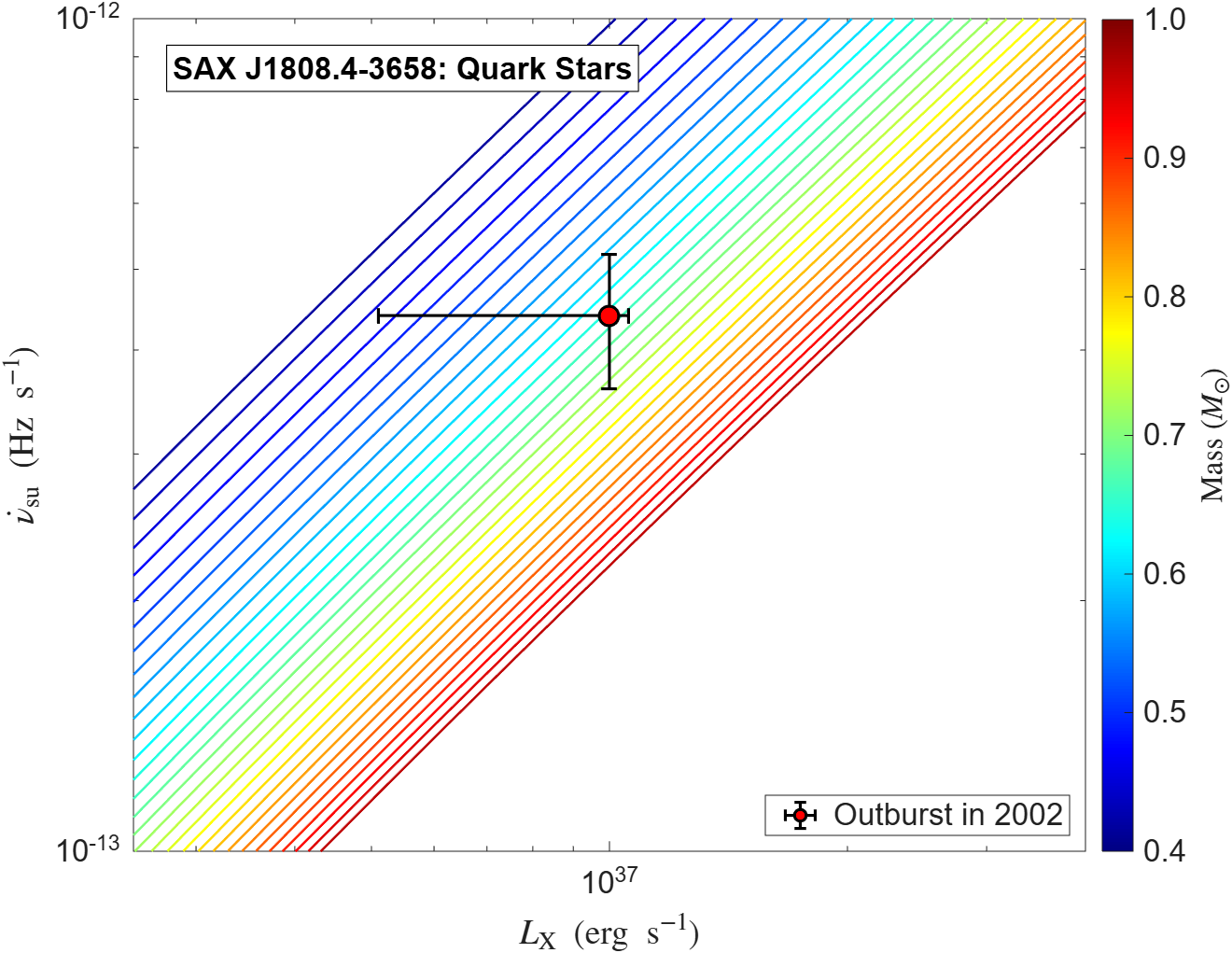}
    \includegraphics[width=0.9\columnwidth]{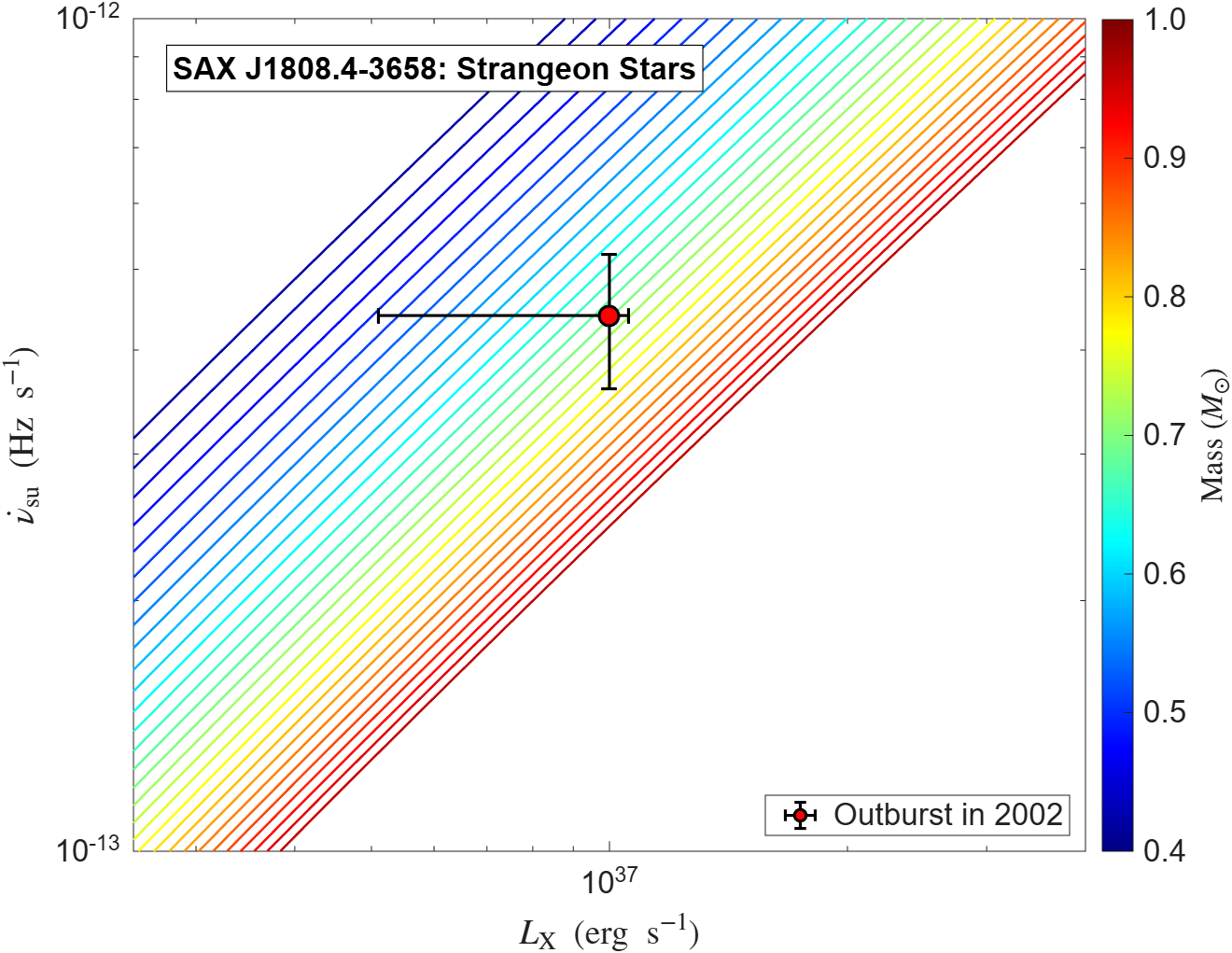}
    \caption{Contours of $M$ with different colors (upper purple one to lower red one: 0.3 to 1 $M_\odot$) for SAX J1808.4-3658, with top, middle and bottom panels representing the cases of neutron stars, quark stars and strangeon stars, respectively. The data point is from the outburst in 2002. The errors in $\dot\nu_{\rm su}$ are from the third column of Table.~\ref{tab:sources}, and the errors in $L_{\rm X}$ are given according to the distance of 2.5-3.6 kpc, with the central point corresponding to 3.5 kpc.  
    }
    \label{fig:results_1808}
\end{figure}

The results for SAX J1808.4-3658 are shown in Fig.~\ref{fig:results_1808}, and from the top to bottom panels are cases of neutron stars, quark stars and strangeon stars, respectively.
In each panel, the contours of $M$ with different colors correspond to masses ranging from 0.4 to 1 $M_\odot$, from upper purple one to lower red one.
The data point is from the outburst in 2002. 
The errors in $\dot\nu_{\rm su}$ are from the third column of Table.~\ref{tab:sources}, and the errors in $L_{\rm X}$ are given according to the distance of 2.5-3.6 kpc~\citep{intZand2001A&A,Galloway2006ApJ}, with the central point corresponding to 3.5 kpc. 
Comparing to the results of XTE J1751–305 in Fig.~\ref{fig:results_1751}, SAX J1808.4-3658 gives substantially smaller masses.   
Although the three EoS models yield slightly different results, the data point falls within a similar range centered around 0.7 $M_\odot$.

\begin{figure}
	\centering
    \includegraphics[width=0.9\columnwidth]{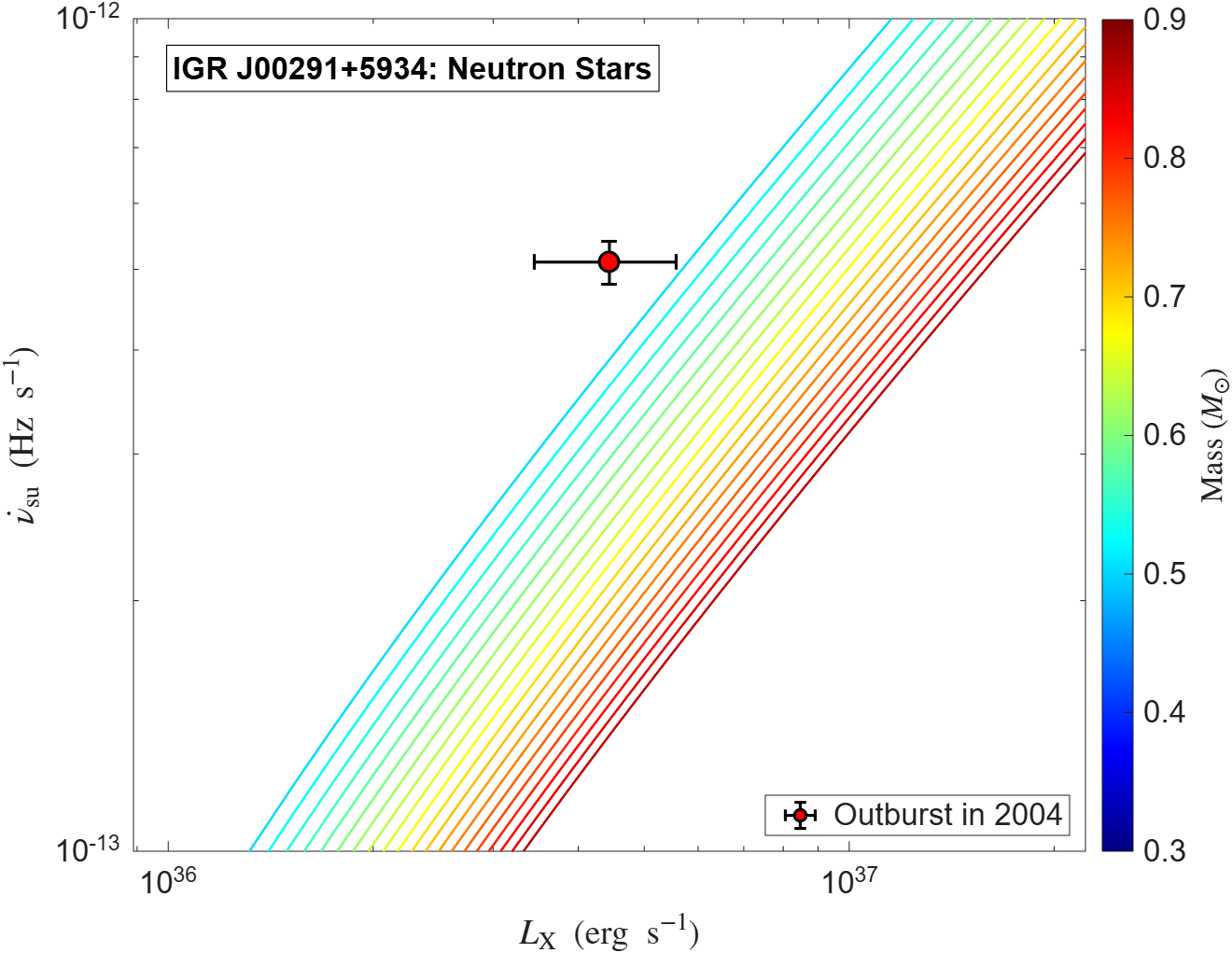}
    \includegraphics[width=0.9\columnwidth]{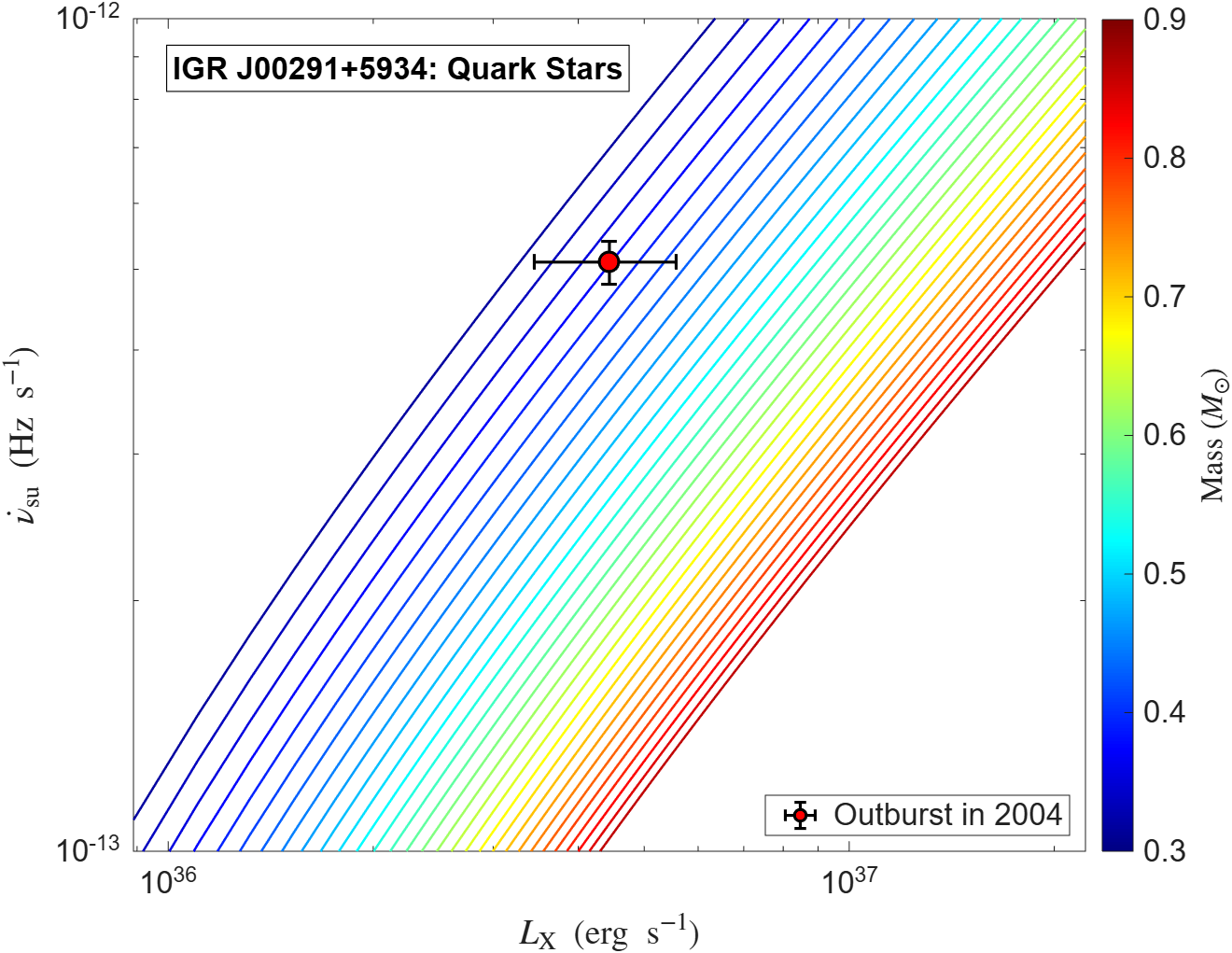}
    \includegraphics[width=0.9\columnwidth]{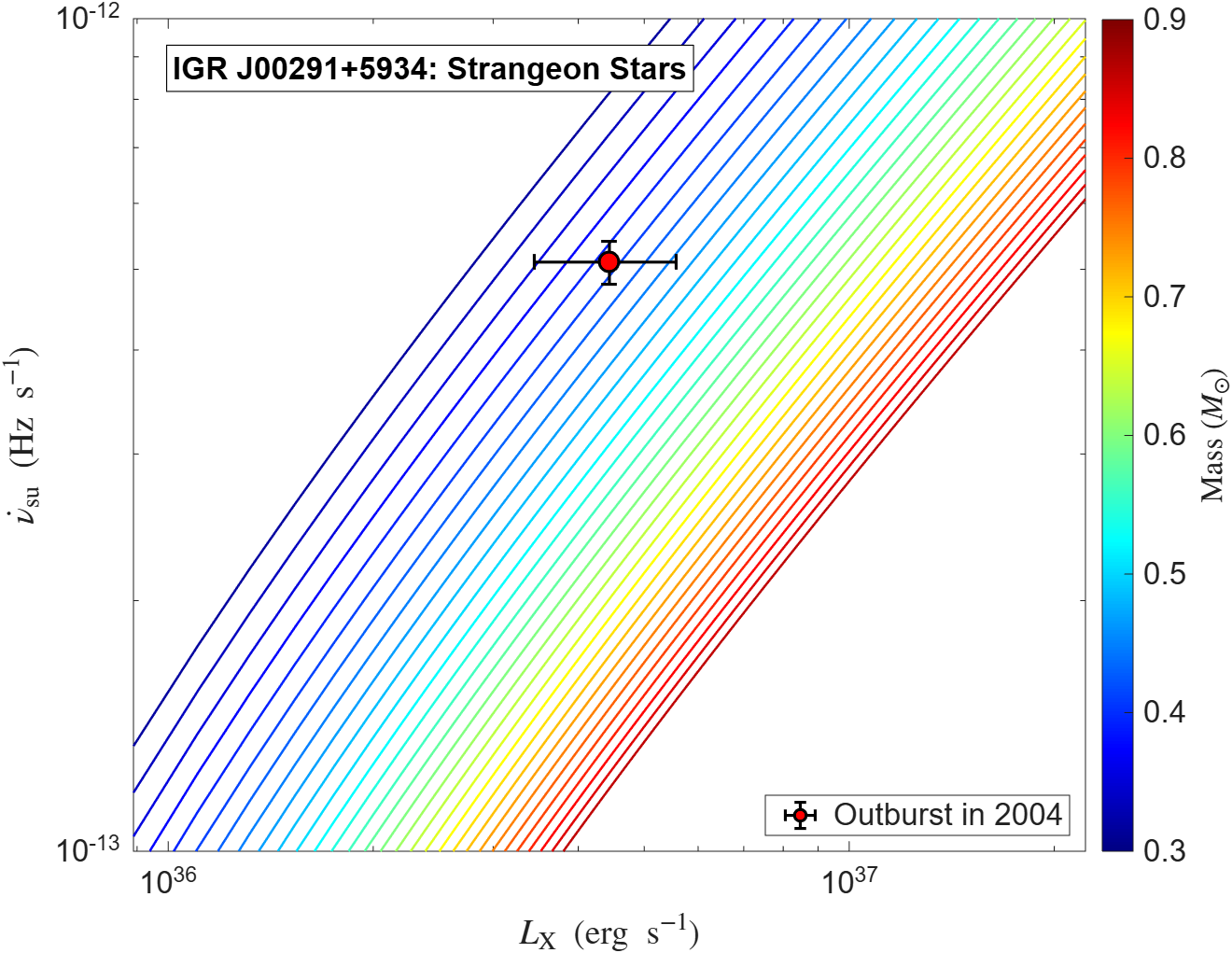}
    \caption{Contours of $M$ with different colors (upper purple one to lower red one: 0.3 to 1 $M_\odot$) for SAX J1808.4-3658, with top, middle and bottom panels representing the cases of neutron stars, quark stars and strangeon stars, respectively. The data point is from the outburst in 2004. The errors in $\dot\nu_{\rm su}$ are from the third column of Table.~\ref{tab:sources}, and the errors in $L_{\rm X}$ are given according to the distance of 4.2$\pm$0.5 kpc.  
    }
    \label{fig:results_00291}
\end{figure}

The results for IGR J00291+5934 are shown in Fig.~\ref{fig:results_00291}, with top, middle and bottom panels representing the cases of neutron stars, quark stars and strangeon stars, respectively.
In each panel, the contours of $M$ with different colors correspond to masses ranging from 0.3 to 0.9 $M_\odot$, from upper purple one to lower red one.
The data point is from the outburst in 2004. 
The errors in $\dot\nu_{\rm su}$ are from the third column of Table.~\ref{tab:sources}, and the errors in $L_{\rm X}$ are given according to the distance of 4.2$\pm$0.5 kpc~\citep{DeFalco2017A&A}. 
Comparing to the results of SAX J1808.4-3658 and XTE J1751–305 in Fig.~\ref{fig:results_1751} and Fig.~\ref{fig:results_1808}, IGR J00291+5934 gives surprisingly small masses.   
For both models of quark star and strangeon star, the data point falls within the range around 0.4 $M_\odot$.
The neutron star model seems inconsistent with the observations, since the mass cannot be that small.

\section{Conclusions and Discussions}
\label{sec:sec_con}

We examine the role of the global structure of AMXPs in accretion-induced spin-up, using structural parameters derived from specific EoS models rather than adopting typical values.
The accretion torque is adopted to be of the classical form written in Eq.(\ref{eq:torque}), and values of $M$, $R$ and $I$ are derived from specific EoS models, including neutron stars (NSs), quark stars (QSs) and strangeon stars (SSs). 
The accretion rate $\dot{M}$ and magnetic field $B$ are derived from Eq.(\ref{eq:Lx}) and Eq.(\ref{eq:B}) respectively, where the EoS-model dependent values of $M$, $R$ and $I$ are also considered. 
Using the observational results of spin frequency $\nu$ and the long-term spin-down rate $\dot{\nu}_{\rm sd}$, we can get the relation between bolometric X-ray luminosity $L_{\rm X}$ and spin-up rate $\dot\nu_{\rm su}$ during outbursts.

The contours of $M$, plotted in the $L_{\rm X}$-$\dot{\nu}_{\rm su}$ plan, could be used to constrain masses of AMXPs by comparing to the observational data of $L_{\rm X}$ and $\dot{\nu}_{\rm su}$ during their outbursts.
The results seem insensitive to EoS models, as the mass constraints by a given AMXP outburst are broadly consistent under the three EoS models. 
Therefore, accretion-induced spin-up could theoretically serve as an evolutionary channel for constraining the values of $M$.

The results show that, the mass of XTE J1751–305 constrained by its 2002 outburst is centered around 1.4 $M_\odot$, the mass of SAX J1808.4-3658 constrained by its 2002 outburst is centered around 0.7 $M_\odot$, and the mass of IGR J00291+5934 constrained by its 2004 outburst is centered around 0.4 $M_\odot$.
The masses of SAX J1808.4-3658 and IGR J00291+5934, especially the latter, are constrained to be very small, since both of them have relatively large values of spin-up rate $\dot\nu_{\rm su}$ while the values of $L_{\rm X}$ are very low.
A low $L_{\rm X}$ implies a low accretion rate, so a large $\dot\nu_{\rm su}$ can be achieved by reducing the mass.

Comparison among results of different EoS models shows that, the NS model yields a slightly larger mass constraint than QS and SS models, for the same outburst source. 
This is because, at a given mass, the radius of the NS is larger than that of the QS and SS (at least for masses below $\sim 1.5\ M_\odot$)~\citep{Gao2022MNRAS}, while the difference in the moment of inertia is comparatively smaller. 
Hence, according to Eq.(\ref{eq:torque}), for the same magnetic field and accretion rate, the NS could experience a larger spin-up rate at a given mass. 
Since a larger mass is less favorable for spin-up, the $M$-contours of NSs shown in Fig.~{\ref{fig:results_1751}} and Fig.~\ref{fig:results_1808} are slightly shifted upward in color compared to those of the QSs and SSs.

The mass constraints from the three EoS models for a given AMXP outburst are broadly consistent, but they depend sensitively on the observed $L_{\rm X}$ and $\dot{\nu}_{\rm su}$.
Taking the 2002 outburst of XTE J1751–305 as an example, a 20\% change in the distance leads to a corresponding change in $L_{\rm X}$, which in turn causes the inferred mass to increase or decrease by about 0.3 $M_\odot$.
Hence, future improvements in the measurements of $L_{\rm X}$ and $\dot{\nu}_{\rm su}$ would refine our results.

The unusually small masses inferred for IGR J00291+5934 and SAX J1808.4-3658, may point to the possible refinements include the following: 
(1) The observed values of $L_{\rm X}$ used in our analysis may be inaccurate, possibly due to distance misestimation. For IGR J00291+5934 to have a mass of 1.4 $M_\odot$, its distance would need to be about 12.5 kpc; for SAX J1808.4-3658 to reach the same mass, a distance of about 6.5 kpc would be required.
(2) The efficiency with which gravitational energy is converted into X-ray luminosity could be significantly below unity. In this scenario, the same $L_{\rm X}$ would imply a larger accretion rate and hence a larger mass.
(3) The adopted accretion torque formulation may require revision. To concentrate on the influence of the stellar structure on the accretion-induced spin-up, here we use the classical form of accretion torque. This certainly could be improved by more detailed modeling considering the complex interaction between the disc and magnetic field lines.

While the results presented here require further refinement on both theoretical and observational grounds, the potential to constrain the values of $M$ remains useful.
In this paper, we find that for SAX J1808.4-3658, the observational mass constraint is below 1 $M_\odot$ --- not as extreme as for IGR J00291+5934, but still outside the generally accepted range for NS model.
The reason is that in NS model it is difficult for pulsars born in supernovae to accommodate such low masses; moreover, the rapid rotation suggests that a non-negligible amount of mass has been accreted, making such a low mass even less plausible. 
Pulsars in the QS model and SS model, in contrast, may naturally have masses below $1\ M_\odot$.
Therefore, if future improvements in the measurements of $L_{\rm X}$ and $\dot{\nu}_{\rm su}$ could provide a more reliable constraint on the mass of an AMXP (e.g., below $1\ M_\odot$), this could offer crucial evidence for probing the nature of pulsar-like compact stars.

The principal innovation of this paper may lie in demonstrating that the use of EoS-specific structural parameters, as opposed to canonical values, could lead to new insights.
We have previously employed this methodology to distinguish different EoS models using the masses and spin periods of recycled MSPs~\citep{ZhongMNRAS2025}.
Extending this framework to other astrophysical contexts involving pulsar-like compact objects --- such as neutron star ultraluminous X-ray sources (ULXs) --- would be a promising direction for future investigations.

\section*{Acknowledgements}

This work is supported by the National SKA Program of China (No. 2020SKA0120300).

\section*{Data Availability}

The observational data used in this work for comparison with our theoretical model are cited under Table.\ref{tab:sources}. No new observational or simulation data were generated in this study. The theoretical model and derived results are fully described in the text. All intermediate data and analysis scripts are available from the corresponding author upon reasonable request.











\bsp	
\label{lastpage}
\end{document}